\documentclass[reprint,longbibliography,superscriptaddress,amsmath,amssymb,aps,prx]{revtex4-2}

\usepackage{amsmath,amssymb,amsthm,mathrsfs,amsfonts,dsfont}
\usepackage{qcircuit}
\usepackage[dvips]{graphicx}
\usepackage{braket}
\usepackage{bm}
\usepackage{bbm}
\usepackage{enumerate}
\usepackage{color}
\usepackage{comment}
\usepackage{hyperref}
\hypersetup{
    colorlinks=true,
    linkcolor=blue,
    filecolor=blue,      
    urlcolor=blue,
    citecolor=blue,
    pdftitle={Efficient Simulation of Leakage Errors in Quantum Error Correcting Codes Using Tensor Network Methods},
    pdfpagemode=FullScreen,
}
\usepackage[normalem]{ulem}

\begin{document}
\title{Efficient Simulation of Leakage Errors in Quantum Error Correcting Codes Using Tensor Network Methods}

\author{Hidetaka Manabe}
\affiliation{Graduate School of Engineering Science, Osaka University,
1-3 Machikaneyama, Toyonaka, Osaka 560-8531, Japan}

\author{Yasunari Suzuki}
\affiliation{NTT Computer and Data Science Laboratories, NTT Corporation, Musashino 180-8585, Japan}
\affiliation{JST, PRESTO, 4-1-8 Honcho, Kawaguchi, Saitama, 332-0012, Japan}

\author{Andrew S. Darmawan}
\affiliation{Yukawa Institute of Theoretical Physics (YITP), Kyoto University,
Kitashirakawa Oiwakecho, Sakyo-ku, Kyoto 606-8502, Japan}
\affiliation{JST, PRESTO, 4-1-8 Honcho, Kawaguchi, Saitama, 332-0012, Japan}

\begin{abstract}
Leakage errors, in which a qubit is excited to a level outside the qubit subspace, represent a significant obstacle in the development of robust quantum computers. We present a computationally efficient simulation methodology for studying leakage errors in quantum error correcting codes (QECCs) using tensor network methods, specifically Matrix Product States (MPS). Our approach enables the simulation of various leakage processes, including thermal noise and coherent errors, without approximations (such as the Pauli twirling approximation) that can lead to errors in the estimation of the logical error rate. We apply our method to two QECCs: the one-dimensional (1D) repetition code and a thin $3\times d$ surface code. By leveraging the small amount of entanglement generated during the error correction process, we are able to study large systems, up to a few hundred qudits, over many code cycles. 
We consider a realistic noise model of leakage relevant to superconducting qubits to evaluate code performance and a variety of leakage removal strategies. Our numerical results suggest that appropriate leakage removal is crucial, especially when the code distance is large.

\end{abstract}
\maketitle

\section{Introduction}
Overcoming the effect of noise is crucial for practical applications of quantum computers. A known method to suppress qubit errors is to use a quantum error correcting code~(QECC)~\cite{shor_Scheme_1995,gottesman_Stabilizer_1997,krinner_Realizing_2022, zhao_Realization_2022, acharya_Suppressing_2023, miao_Overcoming_2022,acharya2024quantum}, in which protected logical qubits are encoded into a larger number of physical qubits, and physical errors can be detected by appropriate measurements. 

One type of error that is not typically corrected by QECCs is leakage error~\cite{aliferis_FaultTolerant_2006,motzoi_Simple_2009}, 
where a qubit escapes from its computational subspace. Recent experimental reports suggest that these errors can severely impact the performance of QECCs~\cite{miao_Overcoming_2022}.
It is not effective to increase the code distance for mitigating this type of error because typical QEC codes are designed to reduce qubit-space errors, not leakage errors.

Furthermore, qubit control pulses are designed to achieve high performance when qubits are in the qubit space. Two-qubit operations acting on leaked qubits may have undesirable effects, such as moving  surrounding qubits to leaked states. Thus, leakage errors are considered one of the major issues in scaling fault-tolerant quantum computing up to practical regimes. So far, massive efforts have been devoted to analyzing the effect of leakage errors and devising methods for their removal~\cite{fowler_Coping_2013,brown_Leakage_2019,varbanov_Leakage_2020,bultink_Protecting_2020,mcewen_Removing_2021,battistel_HardwareEfficient_2021,miao_Overcoming_2022}. 

Simulating the effect of leakage and evaluating the performance of leakage removal strategies accurately on large scale QECCs is challenging since leakage errors are not Pauli errors and therefore cannot be directly incorporated into standard stabilizer simulations~\cite{aaronson_Improved_2004}.
Nevertheless, Ref.~\cite{miao_Overcoming_2022,acharya_Suppressing_2023,acharya2024quantum} used simplified leakage noise models, based on the Pauli twirl approximation, to treat leakage errors with the stabilizer formalism. Yet, it is not clear how well these simplified models capture the performance of realistic leakage in QECCs, since the Pauli approximations discard the coherence term of noise and leads to significantly misrepresented performance for other types of noise~\cite{ darmawan_TensorNetwork_2017,bravyi_Correcting_2018}. Therefore, there is a strong demand for developing scalable and accurate simulation methods to understand how leakage errors degrade the performance of QECCs and how they can be removed.

In this paper, we use tensor-network methods ~\cite{bridgeman_Handwaving_2017,cirac_Matrix_2021,okunishi_Developments_2022} to efficiently simulate QEC codes under leakage errors without approximations that simplify the noise model. We utilize the Matrix Product State (MPS) tensor-network ansatz~\cite{perez-garcia_Matrix_2007} for simulating the one-dimensional (1D) repetition code~\cite{riste_Detecting_2015,kelly_State_2015,suzuki_Efficient_2017,chen_Exponential_2021} and the thin $3\times d$ surface code under leakage errors. This method allows efficient simulation of large numbers of qubits by exploiting the fact that the amount of bipartite entanglement generated by noisy repetitive stabilizer measurements in the QEC process is small enough to be accurately approximated by an MPS. We can simulate a few hundred qutrit systems, a scale unreachable even with the latest state-vector simulators~\cite{suzuki_Qulacs_2021,quantum_ai_team_and_collaborators_2020_4062499,Qiskit,guerreschi_Intel_2020}.

We have used our simulator to study a variety of leakage processes including coherent leakage and thermal noise.
Our numerical results show that there can be a large discrepancy in performance between a coherent noise model and a stochastic approximation to the coherent noise model.
These findings reveal that simplified metrics for leakage, such as leakage and seepage rates, are not entirely sufficient in capturing the real influence of leakage, and
the ability of our method to capture general non-Pauli noise processes can be very useful 
%is essential
for assessing the code performance accurately.

We have also used this method to compare the performance of different leakage removal strategies. 
We found that the effect of leakage removal strategies significantly depends on the details of the leakage process, with some differences becoming distinctly apparent only in regions with large code distances.
These results underscore the necessity of simulating QECCs, particularly in large-scale quantum systems, in order to design qubits that are resistant to leakage.
We expect our methods, which fulfill the requirements of both accuracy and scalability, will indispensable tools for developing scalable quantum computing with practical quantum devices.

The paper is structured as follows. We describe the leakage noise model settings in Sec.~\ref{sec:leakage_noise_modes} and the tensor network simulation methods in Sec.~\ref{sec:tensor_network_methods}. We formulate the problem settings of the QEC codes used in this study and review some leakage removal strategies in Sec.~\ref{sec:problem_setting}. In Sec.~\ref{sec:result}, we present the main numerical results on the MPS simulations of the 1D repetition code and the $3\times d$ thin surface code and discuss the effect of leakage. We conclude the paper by summarizing our contribution and future directions in Sec.~\ref{sec:conclusion}.

\section{Leakage noise models}\label{sec:leakage_noise_modes}
We aim to provide a general simulation framework that can efficiently evaluate the performance of QECCs under leakage error models and compare the mitigation methods for them. In this section, we introduce several phenomenological leakage noise models to demonstrate our simulation framework.

Noise, including leakage noise, may occur during any operations such as gates, measurements, or when qubits are idle, {and they are observed in a variety of qubit devices, such as superconducting circuits~\cite{acharya2024quantum}, neutral atoms~\cite{cong2022hardware} and trapped ions~\cite{hayes2020eliminating}}. Their noise modeling highly depends on each qubit system and architecture. In this paper, we focused on superconducting qubits~\cite{nakamura_Coherent_1999,koch_Chargeinsensitive_2007,barends_Superconducting_2014,kjaergaard_Superconducting_2020} as an example target of noise modeling. {This is because they achieved the demonstration of surface codes for up to $d=7$ and faced the problems of leakage errors~\cite{acharya_Suppressing_2023,acharya2024quantum}.}
To concisely demonstrate the effect of typical leakage errors on QECCs, we assume superconducting qubits as a qutrit system and employ the leakage noise models discussed below. 
While these noise models are not intended to correspond to any specific experimental noise process, they nevertheless capture key features of leakage, such as coherence and leakage spreading. More detailed noise models, corresponding to specific experimental architectures, can be straightforwardly taken into account using our method, albeit potentially at a higher computational cost. 
%Note that our simulation methods can also be used for regenerating experimental results with some additional computational cost by using more faithful noise models for specific architecture and considering higher levels of physical systems. Nevertheless, we employ this simplification to observe the qualitative behavior of QECCs to leakage errors. The faithful regeneration of experimental results is left as future work.
%Note that our simulation method could include more leakage levels, and we probably need more leakage levels to regenerate reported experimental results, 

\subsection{Single-qubit gate errors}
A significant factor that incurs leakage in single-qubit operation is the use of short pulses for fast control~\cite{motzoi_Simple_2009}.
%While short pulse controls have the advantage that they reduce the probability of qubit decay during a gate, the disadvantage of short pulses is that they have a broad frequency spectrum and may excite qubits to the leakage space.
%Since the anharmonicity of transmon qubits is typically designed to be small, this effect is non-negligible, and thus there is a trade-off relation between errors from lifetime decay and fast-control leakage.
The exact modeling of the leaking unitary operator is analytically difficult and is highly dependent on specific details of the architecture. 
To simplify the noise model, we make the following assumptions. 1) Rabi-oscillation in the qubit space and leakage states happens independently. 2) The oscillations between $\ket{0}\leftrightarrow \ket{2}$ and $\ket{1}\leftrightarrow \ket{2}$ can be treated independently. 3) The oscillation between leaked states can be modeled as far-detuned Rabi oscillation.
According to the above assumptions, we account for the coherent transfer between the qubit space and the leaked state by fast control as follows.
\begin{equation}
    U=U_{z}(\varphi_i)\cdot R_{02}(\theta,\lambda_i)\cdot R_{12}(\theta,\lambda_i)
\end{equation}
where
\begin{align}
    &R_{02}(\theta,\lambda_i)=U_{02}(\theta,\lambda_i)\oplus I_1 \\
    &R_{12}(\theta,\lambda_i)=U_{12}(\theta,\lambda_i)\oplus I_0 \\
    &U(\theta,\lambda_i)=e^{-i\theta/2}\exp(i\theta/2(\cos\lambda_i X+\sin\lambda_i Y))
\end{align}
and
\begin{align}
    &U_z(\varphi_i)=\begin{pmatrix}
        1 & 0 & 0 \\
        0 & 1 & 0 \\
        0 & 0 & e^{i\varphi_i} \\
    \end{pmatrix}
\end{align}
Here, $\theta$ denotes the strength of rotation noise, while $\lambda_i$ and $\varphi_i$ denote the relative phases of leakage rotations. While in experiment these phases are determined by the details of physical systems, in this study they are randomly chosen from $[0, 2\pi)$ for each qubit. $I_0$ and $I_1$ are identity operators on the 1D subspaces $\{\ket0\}$ and $\{\ket1\}$, respectively. We model the occurrence of the following unitary rotation after applying the CZ gate.

\subsection{Controlled-phase gate errors}
Two-qubit operations applied to leaked qubits can have undesired effects. 
We consider two effects from applying two-qubit gates acting on leaked qubits: undesired phase shifts, and leakage spreading.
They have been experimentally observed in Ref.~\cite{miao_Overcoming_2022} and reported as major noise factors that increase logical error rates in Ref.\,\cite{acharya2024quantum}. 
To model this, we define the noisy CZ gate as follows:
\begin{align}
\mathrm{CZ_{noisy}}=\begin{pmatrix}
1 & 0 & 0 & 0 & 0 & 0 & 0 & 0 & 0 \\
0 & 1 & 0 & 0 & 0 & 0 & 0 & 0 & 0 \\
0 & 0 & 1 & 0 & 0 & 0 & 0 & 0 & 0 \\
0 & 0 & 0 & 1 & 0 & 0 & 0 & 0 & 0 \\
0 & 0 & 0 & 0 &-1 & 0 & 0 & 0 & 0 \\
0 & 0 & 0 & 0 & 0 & i & 0 & 0 & 0 \\
0 & 0 & 0 & 0 & 0 & 0 & 1 & 0 & 0 \\
0 & 0 & 0 & 0 & 0 & 0 & 0 & i & 0 \\
0 & 0 & 0 & 0 & 0 & 0 & 0 & 0 & 1 \\
\end{pmatrix}
\end{align}
and we define the noisy CNOT gate as $(I\otimes H)\mathrm{CZ_{noisy}}(I\otimes H)$ where
\begin{align}
    H=\begin{pmatrix}
        \frac{1}{\sqrt{2}} & \frac{1}{\sqrt{2}} & 0 \\
         \frac{1}{\sqrt{2}}& \frac{-1}{\sqrt{2}} & 0\\
        0 & 0 & 1
    \end{pmatrix}\,.
\end{align}
When both systems are in the qubit space, the noisy CZ and CNOT gates act as the ideal versions of these gates. 
However, a phase of $i$ will be added when a control qubit is in the leaked space.
{Here, we chose phase $i$ ($\pi/2$ phase shift) for CZ gates on $\ket{2}$ state since, according to Ref.\,\cite{miao_Overcoming_2022}, the observed phase shift for this state is intermediate of $\ket{0}$ state ($0$ phase shift) and $\ket{1}$ state ($\pi$ phase shift).}

Leakage spreading has also been observed in experiment~\cite{miao_Overcoming_2022}, in which CZ gates on qubits in a leaked state cause neighboring qubits to leak. This phenomenon is understood as a composition of multiple transitions between states with higher levels~\cite{mcewen_Removing_2021}. In this study, we consider a simpler phenomenological model of leakage spreading consisting of a sequence of rotations about the Y axis in the subspaces $\{\ket{02},\ket{22}\},\{\ket{12},\ket{22}\},\{\ket{20},\ket{22}\},\{\ket{21},\ket{22}\}$ with angle $\theta_{\mathrm{spread}}$. These four operations are applied after the CZ gate and cause a state in the qubit subspace to coherently rotate into the leakage subspace when the neighboring qubit is in the leaked state.

\subsection{Measurement}
%Superconducting qubits are typically measured with a dispersive readout scheme. In this scheme, measurement results are obtained by discriminating response microwave pulses to 0 or 1. In our noise model, we assume that, when a qubit is in the $\ket2$ state, the response pulse is randomly classified to 0 or 1.
Typical measurement schemes, such as dispersive readout in superconducting qubits, are not designed to discriminate leaked states from the qubit space. Thus, we assume that when a qutrit is in the leaked state, the measurement results are randomly classified as 0 or 1 with 1/2 probability.
%We can simulate this effect by performing the perfect projection measurement in $\{\ket0, \ket1, \ket2\}$ and, if the result is 2, we randomly replace the value with $0$ or $1$. 
With this assumption, the action of measurements can be modeled using the following CP-instrument:
\begin{align}
    &\mathcal{E}_0(\rho)=\Pi_0\rho\Pi_0+p\Pi_2\rho\Pi_2 \\
    &\mathcal{E}_1(\rho)=\Pi_1\rho\Pi_1+(1-p)\Pi_2\rho\Pi_2
\end{align}
where $p=0.5$ and $\Pi_i=\ket{i}\bra{i}$.
%where $p\in[0, 1]$ and $\Pi_i=\ket{i}\bra{i}$.
%For simplicity, we set $p=0.5$ in our numerical evaluation.

{Measurements can also be a significant contributor to leakage errors~\cite{sank2016measurement,shillito2022dynamics,khezri2023measurement}, and can easily be taken into account with a tensor network simulator (they can simply be absorbed into gate leakage before the measurement). However, for simplicity, in this work, we neglect measurement leakage and instead focus on the dominant sources of leakage, namely CZ and thermal leakage~\cite{acharya2024quantum}.}

\subsection{Idling noise}
Even when qubits are idle, the thermal noise, i.e., amplitude damping and thermal excitation, may excite a qubit state to out-of-qubit states. {This effect is typically not a dominant error source but is reported as a non-negligible contributor to leakage errors in TABLE~S4 in Ref.\,\cite{acharya2024quantum}.}
%The effect of thermal noise can be modeled by considering qubits to be coupled to finite-temperature heat baths. Thus, when we take too long for qubit control, a finite fraction of the population is moved to higher-energy states. 
%This effect is negligible for atoms since the transition energy is much higher than the room temperature. However, this is not negligible in the case of microwave qubits such as quantum dots and superconducting qubits.
For formulating the thermal noise model, for simplicity, we utilize the dynamics of harmonic oscillators connected to thermal baths written as:
\begin{align}
    \begin{split}
        \dot\rho = \gamma(N+1)(a\rho a^\dagger-\frac{1}{2}\{a^\dagger a,\rho\})\\
            +\gamma N(a^\dagger\rho a-\frac{1}{2}\{aa^\dagger,\rho\}),
    \end{split}
    \label{eq:ADTE}
\end{align}
where $N = (e^{\hbar\omega/k_BT}-1)^{-1}$, $a$ and $a^{\dagger}$ are annihilation and creation operators, $\gamma$ is the coupling strength to the bath, and $k_BT/\hbar\omega$ is the effective temperature relative to the photon energy. 
%Assuming the resonant frequency of the qubit is $10 \mathrm{GHz}$, we can rewrite $k_BT/\hbar\omega:=\alpha T$ and $\alpha\sim 13.1\,\mathrm{K}^{-1}$. 
We calculate the CPTP-map with Kraus representation of the above model by cutting off the higher levels and integrating the above dynamics over a period of gate time. 
We apply this error map to all qubits at the beginning of each syndrome measurement round.
In the simulation, we set the resonant frequency of qubit as $10 \mathrm{GHz}$, which means $k_BT/\hbar\omega:=\alpha T$ and $\alpha\sim 13.1\,\mathrm{K}^{-1}$. We assume $\tau=1.0\,\mu s$ and units of $\gamma$ and $T$ being $\mathrm{MHz}$ and $\mathrm{mK}$, respectively for the numerical calculations of this study.

\section{Tensor network methods}\label{sec:tensor_network_methods}
\begin{figure*}[t]
    \centering
    \includegraphics[width=\linewidth]{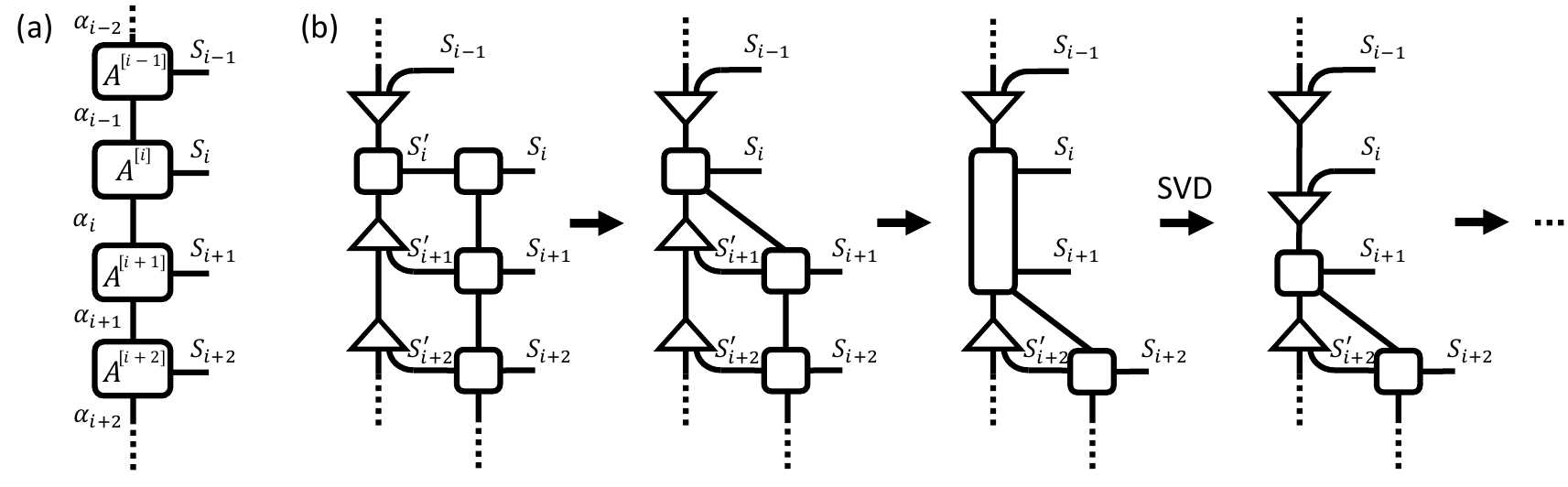}
    \caption{(a) The diagram notation of MPS, a simple tensor network with a 1D structure. (b) The procedure of canonical update. The triangular tensor represents an isometry from two indices (connected to the base of the triangle) to one (connected to the vertex of the triangle), and the square tensor represents the `top tensor', which has no particular structure. When applying a two-qubit gate, the MPS is transformed into a canonical form where the top tensor is located at the $i$-th position, where $i$ is the first site on which the gate acts non-trivially, and the remaining tensors are isometries as pictured. The MPO representing the two-qubit gate is sequentially applied to the MPS from the site $i$. During the process, SVD is used to truncate the bond dimension and restore the state to the canonical MPS format.}
    \label{fig:mps}
\end{figure*}
The noise models described in the previous section include non-Pauli processes such as control leakage, which cannot be efficiently simulated with a Clifford simulator. 
While it is possible to use twirling approximations~\cite{acharya_Suppressing_2023} to convert a non-Pauli noise to a stochastic Pauli one, the approximate noise model is not guaranteed to faithfully reproduce the behavior of the original noise model.

In order to simulate general circuits, state-vector~\cite{deraedt_Massively_2007,haner_Petabyte_2017,suzuki_Qulacs_2021} and tensor-network methods~\cite{vidal_Efficient_2003,markov_Simulating_2008,guo_GeneralPurpose_2019,zhou_What_2020,gray_Hyperoptimized_2021,yong_Closing_2021,pan_Solving_2021,huang_Efficient_2021,oh_Tensor_2023,kim_Evidence_2023,tindall_Efficient_2023} are widely used. State-vector type simulations are straightforward to implement but require exponentially large memory capacity: for example, a few petabytes of memory for $30$ qutrit systems.

Tensor-network methods, in contrast, are more complicated but can be used to simulate larger circuits than state-vector approaches. Broadly speaking, tensor-network methods for quantum circuit simulations can be classified into two types: 1) tensor-contraction type~\cite{markov_Simulating_2008,gray_Hyperoptimized_2021,yong_Closing_2021,pan_Solving_2021,huang_Efficient_2021} and 2) ansatz type~\cite{vidal_Efficient_2003,guo_GeneralPurpose_2019,zhou_What_2020,oh_Tensor_2023,kim_Evidence_2023,tindall_Efficient_2023}. 
For the tensor-contraction type, the entire quantum circuit is first converted into a tensor network, simplified, and a near-optimal contraction path is searched heuristically. 
%However, this type of method is not suitable if the intermediate measurements and classical feedback frequently happen.
On the other hand, the ansatz-type method expresses the quantum state during the execution of a quantum circuit by a tensor network with a specific structure. It is known that many interesting quantum states can be efficiently represented using a tensor-network ansatz, which exploits their entanglement structure~\cite{perez-garcia_Matrix_2007,verstraete_Renormalization_2004,vidal_Entanglement_2007,bridgeman_Handwaving_2017,cirac_Matrix_2021,okunishi_Developments_2022}. In this work, we adopt ansatz-type approach and exploit the fact that qubits in many quantum error correcting codes, unlike in generic quantum circuits, have a specific entanglement structure that can be well captured with a tensor-network ansatz. This allows far more efficient simulation compared to tensor-contraction approaches designed for generic circuits. An example of this was seen in previous work \cite{darmawan_TensorNetwork_2017, darmawan_Lineartime_2018},  where the simple tensor-network structure of the surface code was used to probe the effect of non-Pauli noise on error correction performance. However these simulations, unlike those in this work, were limited to the case with noiseless syndrome measurements. 

%This approach is especially useful for simulating medium-size NISQ circuits because they usually generate a large amount of entanglement during the process, which makes the appropriate tensor network approximation difficult. Moreover, in typical NISQ circuits, since we only measure qubits at the final step, we can consider the problem of quantum circuit simulation just as contracting a single large tensor network. 

In this section, we will briefly review the quantum circuit simulation method using Matrix Product States (MPS)~\cite{vidal_Efficient_2003,vidal_Efficient_2004,white_RealTime_2004,zhou_What_2020,noh_Efficient_2020,oh_Tensor_2023,kim_Evidence_2023}, which we use to simulate QECCs.

\subsection{Matrix-product states}

A Matrix Product State~(MPS)~\cite{perez-garcia_Matrix_2007} is a tensor network ansatz typically used to represent 1D pure states. We illustrate the diagram notation of MPS in Fig.~\ref{fig:mps}(a), where nodes correspond to multi-dimensional arrays and each incident edge of a node corresponds to an index of that array. An edge connecting a pair of nodes corresponds to taking sums over corresponding indices. An MPS can be written as
\begin{equation}
\ket{\psi_{s_1,s_2,\cdots s_n}} = \sum_{\{s_i\}}\sum_{\{\alpha_i\}}\left[A_{\alpha_1}^{[1]s_1}A_{\alpha_1\alpha_2}^{[2]s_2}\cdots A_{\alpha_{n-1}}^{[n]s_n}\right]\ket{s_1s_2\cdots s_n}
\end{equation}
where $A^{[i]}$ represent $i$th tensor in Fig.~\ref{fig:mps}(a), and $\{s_i\}$ and $\{\alpha_i\}$ are referred to as physical and virtual indices, respectively. Each amplitude $\braket{s_1s_2\cdots s_n|\psi_{s_1,s_2,\cdots s_n}}$ is expressed by the product of $n$ matrices $\{A^{[i]s_i}\}$.

The edges linking the nodes are called {\it bonds}. The dimension of these bonds is referred to as {\it bond dimension} $\chi$. The expressibility of MPS is constrained by its 1D structure: an MPS with bond dimension $\chi$ can represent quantum states with bipartite entanglement entropy up to $\log_2\chi$ with respect to any bipartition of the chain. 

While MPS is a 1D ansatz, it is often applied to study two-dimensional (2D) systems by snaking the MPS through a 2D lattice e.g. \cite{iregui_Infinite_2017,kim_Evidence_2023}
. While the computational cost will typically grow exponentially with increasing the second dimension, the results of such methods are often competitive with other state-of-the-art methods in 2D. 

\subsection{Quantum circuit simulation with MPS}
Here we briefly describe how the evolution of the many-qubit quantum state over many rounds of quantum error correction can be captured in the MPS formalism.

In the simulation, the state of the quantum system is represented as an MPS, and individual tensors in the MPS are modified according to the action of gates and measurements. Two-qubit gates cause the bond dimension of the MPS to increase. In order to prevent exponential growth in bond-dimension,  it is necessary to perform bond truncation, which we have achieved using a widely known method for MPS called {\it canonical update}. The diagram notation of the canonical update is illustrated in Fig.~\ref{fig:mps}(b).
The MPS is initially put in {\it canonical form}, where all tensors are isometric maps from two indices to one except for one tensor, called the `top tensor'. Any MPS can be put in such a form, where the top tensor can be chosen to be at any position, by performing successive SVDs on individual tensors. When applying a gate, the top tensor is chosen to be the first site on which the gate acts non-trivially. The MPO representing the gate is then applied, according to Fig.~\ref{fig:mps}, followed by an SVD. By truncating small singular values at this stage, the action of the gate and bond truncation are both realized on the MPS. We provide full details of this procedure in Appendix~\ref{app:Quantum_ciricuit_simulation_with_MPS}.

The bond truncation reduces the bond dimension of an MPS to some chosen value $\chi$. In general this will cause some loss in fidelity with the original state $\ket{\psi}$, however the fidelity can be improved systematically by choosing larger values of $\chi$. The value of $\chi$ required to obtain a high fidelity will depend on the entanglement in $\ket{\psi}$. In the error correcting codes described below, we demonstrate that only a small $\chi$ (and therefore small computational cost) is required to represent the state with high fidelity.

\section{Error-correction simulations}\label{sec:problem_setting}
\begin{figure}[t]
    \centering
    \includegraphics[width=\linewidth]{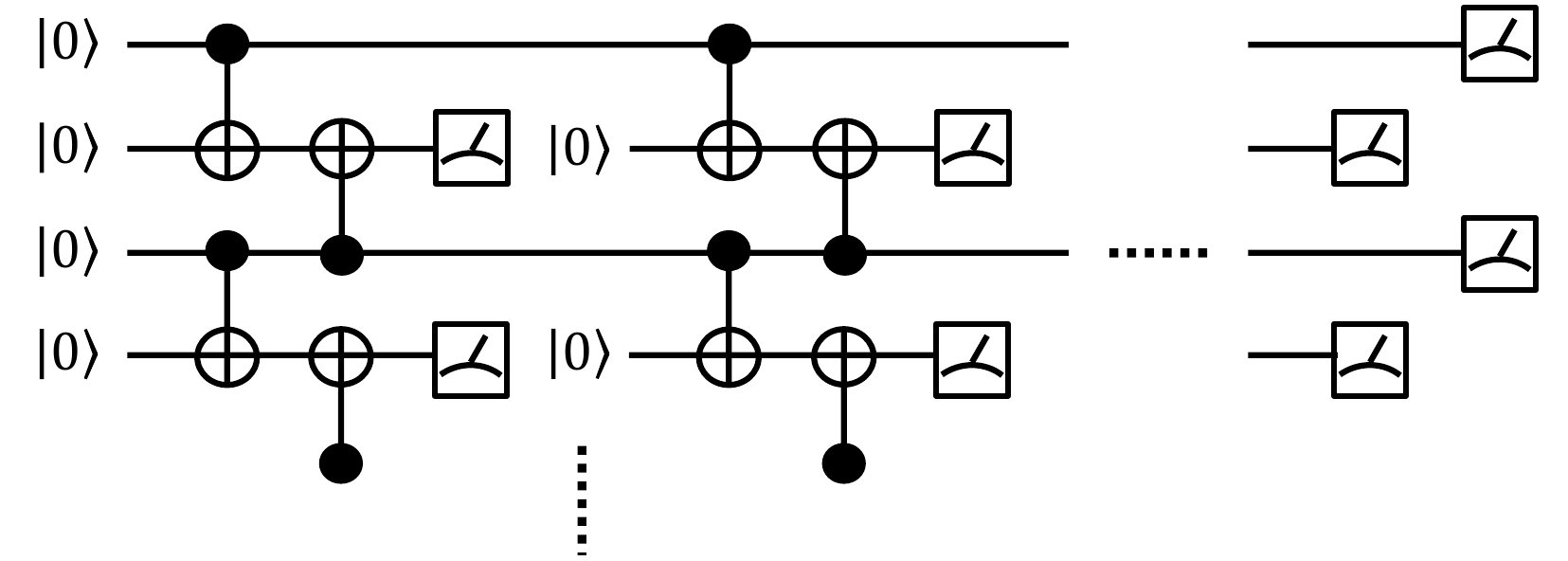}
    \caption{The quantum circuit for the 1D repetition code. Each check is executed by coupling two data qubits to a single ancilla qubit then measuring the ancilla, and repeated $d$ times. At the end, to read out the encoded information, all data qubits are measured. }
    \label{fig:repetition_circuit}
\end{figure}
In this section, we detail the simulation setup to investigate the impact of leakage on QEC codes. We focus on simulations of the 1D repetition code~\cite{riste_Detecting_2015,kelly_State_2015, chen_Exponential_2021} and the thin surface code, which are regarded as simple to implement due to their low connectivity and have been shown theoretically to be highly efficient at correcting biased noise~\cite{tuckett_Tailoring_2019,guillaud_Repetition_2019,darmawan_Practical_2021,ataides_XZZX_2021,higgott_Fragile_2022}. Their respective 1D and quasi-1D entanglement structure makes them ideal targets for the MPS method, described in the previous section. Extending these methods to codes with more complex connectivity in two or higher dimensions will likely require tensor network ansatz beyond MPS, which we leave to future work. 

In addition to standard error correction methods, we also incorporate several leakage removal strategies into the simulation. This allows us to understand the effectiveness of these strategies in minimizing the effect of different types of leakage processes.

\subsection{Error correcting codes}
The quantum circuit of the 1D repetition code is shown in Fig.~\ref{fig:repetition_circuit}.
The logical $\ket0$ and $\ket1$ states for the 1D repetition code are defined as $\{\ket{0\cdots0},\ket{1\cdots1}\}$. We initialize an encoded logical $\ket0$ or $\ket1$ simply by preparing a product of $d$ physical $\ket0$ or $\ket1$ states, respectively at data qubits. 
%We encode a bit into the corresponding logical state using $d$ qubits. 
Then, syndrome measurements are performed $d$ times. Finally, the encoded bit is decoded from the data qubits and syndrome outcomes using the minimum-weight perfect matching algorithm~\cite{dennis_Topological_2002,kelly_State_2015}. Since noise is introduced to the quantum state, decoding does not always succeed. We define the probability of decoding failure as the logical error rate.

\begin{figure}[t]
    \centering
    \includegraphics[width=\linewidth]{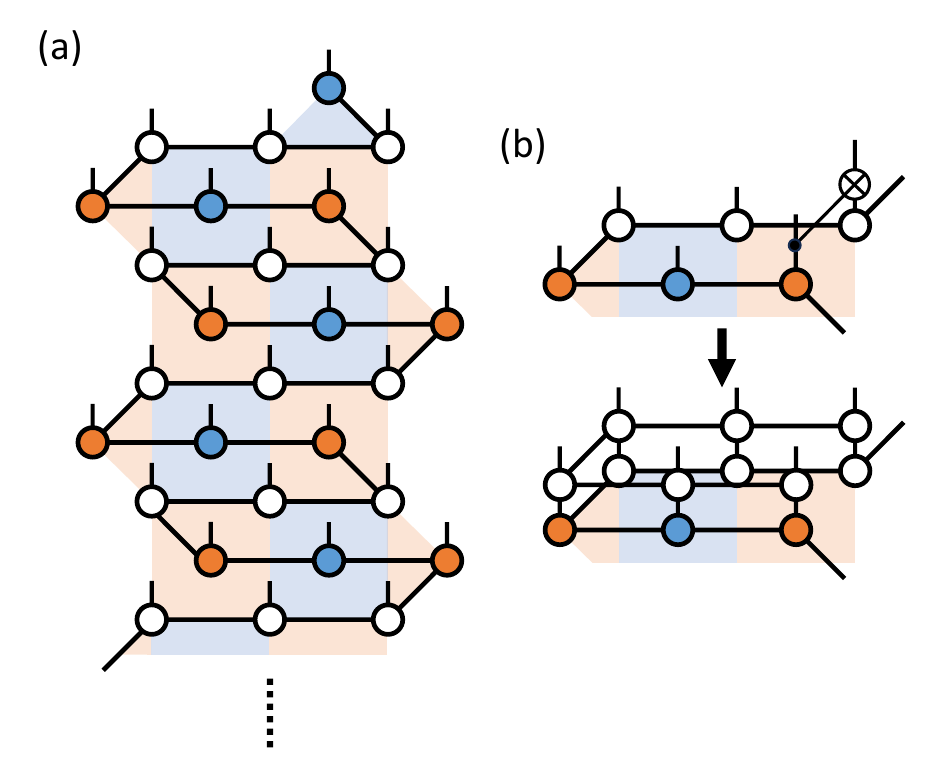}
    \caption{(a) The layout of the thin surface code and corresponding MPS. White vertices represent data qubits, and orange and blue vertices represent ancilla qubits for X-type and Z-type stabilizer measurements, respectively. The MPS is defined in a manner that connects data qubits and ancilla qubits in a snake-like style. (b) Two-qubit gate in the syndrome measurements is represented as MPO.}
    \label{fig:thin_surface}
\end{figure}
We conducted further simulations of the $3\times d$ thin surface code, which has a quasi-1D structure. By thin surface code, we mean a surface code for which the size of the lattice can increase, while its width remains constant. The layout of the thin surface code is illustrated in Fig.~\ref{fig:thin_surface}(a). 
The thin surface codes are considered useful when the noise is biased~\cite{tuckett_Tailoring_2019,darmawan_Practical_2021,higgott_Fragile_2022}. We can reduce the probability of logical $Z$ errors (the vertical $Z$ chain) by increasing $d$. Conversely, the minimum length of the $X$ logical operator is fixed and, as a result, it is anticipated that the logical $X$ error rate will increase with increasing $d$, rather than decrease, due to the larger number of possible horizontal error chains. From this point, we refer to $d$ as the $Z$-distance, corresponding to the minimum number of physical $Z$ errors needed to create a logical error.

In our simulation, both data qubits and ancillary qubits are included together in the tensor-network description of the state. For the 1D repetition code, the data qubits and ancilla qubits are alternately arranged in one dimension, constituting a $2d-1$ qubit system in total if the code distance is $d$. The logical $\ket0$ and $\ket1$ state, including ancilla qubits, is just a product state ($\ket{00\cdots0}$ or $\ket{10101\cdots 1}$ respectively), which can be represented as MPS with $\chi=1$. 
In the case of the $3\times d$ surface code, there are $6d-1$ qubits in total, and we design the MPS structure in a snake-like manner as shown in Fig.~\ref{fig:thin_surface}(a). When the logical $\ket+$ state is encoded, the system is a low-entanglement state, which can also be efficiently represented by MPS with a maximum bond dimension of $\chi=4$. For a 1D repetition code, the two-qubit gate is applied only to neighboring qubits. However, for a thin surface code where the MPS is snake-like, it is necessary to apply an MPO of up to length 6 as shown in Fig.~\ref{fig:thin_surface}(b), escalating the simulation cost.

If there is no noise, the logical state of both codes is a low-entangled state and can thus be efficiently described using MPS. When the noise contains Pauli and a certain simple type of incoherent leakage error, it can also be easily simulated using an extended Clifford simulator, such as Pauli+~\cite{acharya_Suppressing_2023}. When there is coherent noise, if the noise is sufficiently weak and the correlation between errors is small, it is expected that the required bond dimension to reach a certain fidelity will be sufficiently small. This is because the logical state itself is a low-entangled state, and the repetitive stabilizer measurements act to return the entire system to such a state (or a state where a local operator has been applied). This behavior is similar to the measurement-induced phase transitions in monitored quantum circuits~\cite{skinner_MeasurementInduced_2019,bao_Theory_2020}, 
but they differ in that we utilize the properties of specific error-correcting codes and noise models.

We should note that, besides the codes mentioned above, our method can also be applied to various larger QECCs. For the $5\times d$ and $7\times d$ surface codes, for example, only bond dimensions $\chi=8,16$, respectively, are needed at most to represent the logical states, and thus they can be efficiently simulated in low noise regions. However, since the length of the MPO for the 2-qubit gates is long ($10, 14$ respectively), the effect of noise spreads over numerous tensors, resulting in an increase in the required computational resources.

\subsection{Leakage removal strategies}
Leakage noise is known to harm the performance of quantum error correction. Therefore, several strategies have been devised to remove leakage errors~\cite{miao_Overcoming_2022}. In this study, we simulate three approaches to compare and analyze the performance of these strategies.

\subsubsection{No reset}
{\it No reset} strategy does not offer any leakage removal operation after the measurements. In order to set the ancillary qubit to $\ket0$, we only apply the $X$ gate if the outcome is $1$. The quantum channel of this process can be represented as follows
\begin{equation}
    \mathcal{M}_{\mathrm{noreset}}(\rho)=(\bra0\rho\ket0 + \bra1\rho\ket1) \ket0\bra0 + \bra2\rho\ket2\ \ket2\bra2.
\end{equation}

\subsubsection{Multi-level reset}
The {\it multi-level reset (MLR)} strategy involves applying a map to ancillary qubits (and not data qubits) after every syndrome measurement that converts all states including the leakage state perfectly to the $\ket0$ state. This can be expressed as the simple map $\mathcal{M}_{\mathrm{MLR}}(\rho)=\ket0\bra0$.

\subsubsection{DQLR}
\begin{figure}[t]
    \centering
    \includegraphics[width=\linewidth]{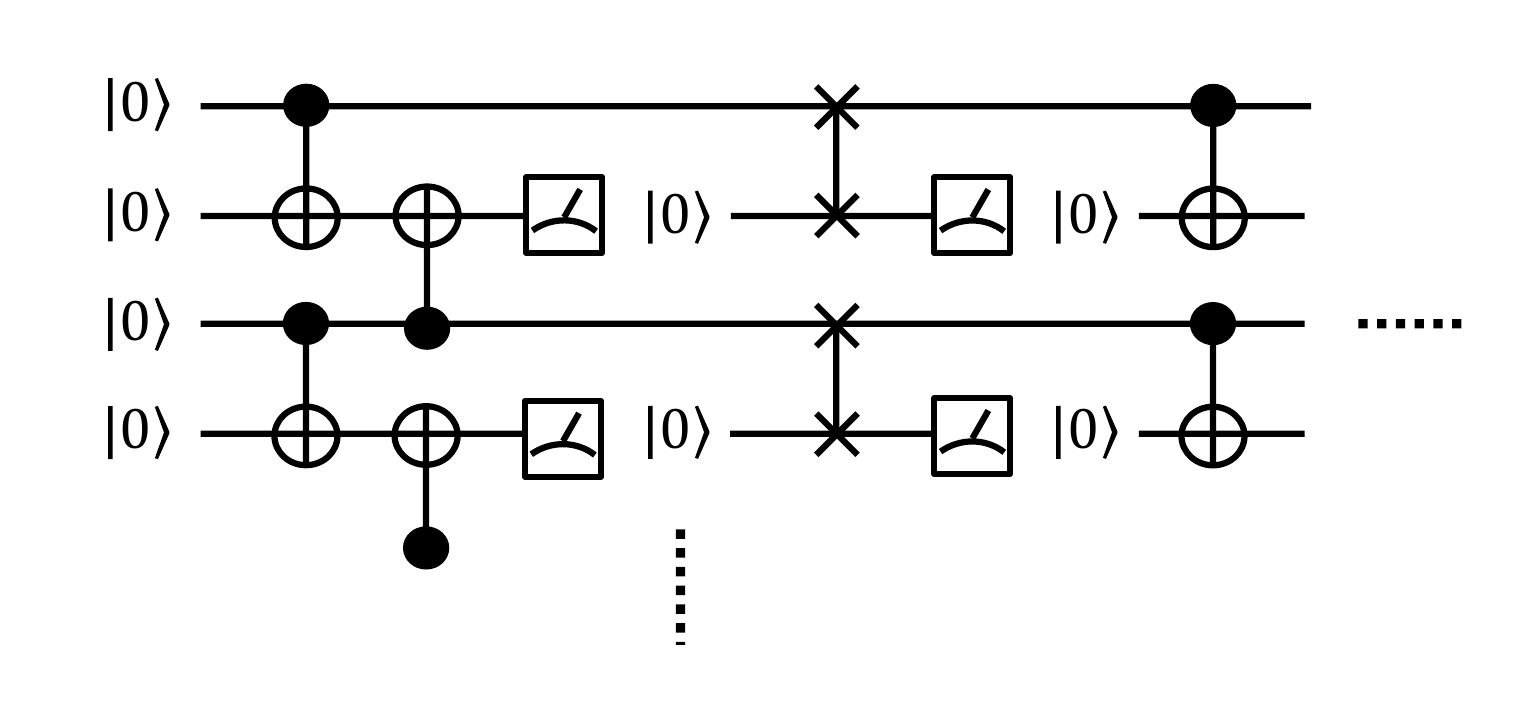}
    \caption{The quantum circuit of the DQLR process. After the stabilizer measurement, the LeakageISWAP gate is applied between pairs of data and ancillary qubits, then ancilla qubits are measured again.}
    \label{fig:DQLR}
\end{figure}
\begin{figure*}[t]
    \centering
    \includegraphics[width=0.9\textwidth]{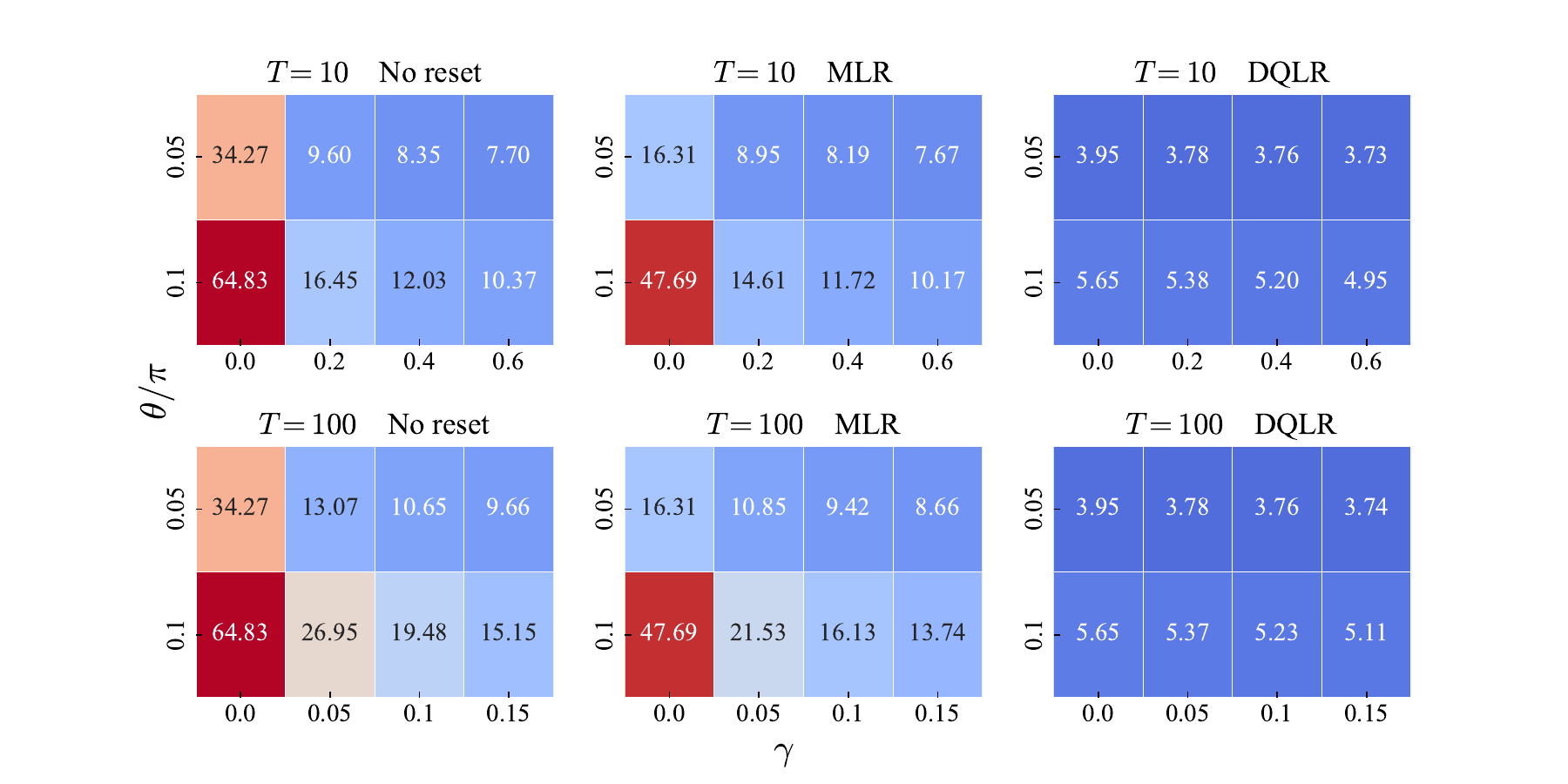}
    \caption{The average MPS bond dimension required to reach a truncation error of $10^{-6}$ when simulating the 1D repetition code for varying the temperature $T$, leakage rotation angle $\theta$, coupling strength to bath $\gamma$, and the leakage removal strategy. The average MPS bond dimension is computed by collecting all the bond dimensions at the end of each round and taking their average. The code distance is $d=99$, the number of rounds is also $99$ and the angle of leakage spreading is fixed as $\theta_{\mathrm{spread}}=0.3\pi$.}
    \label{fig:bond_dim_ave}
\end{figure*}
Although the MLR strategy can remove all leakage in ancillary qubits, it cannot eliminate leakage errors that occur in data qubits, which can cause serious time-correlated errors in QEC circuits. Therefore, in the {\it DQLR} strategy, we apply a $LeakageISWAP$ gate~\cite{miao_Overcoming_2022} between the data qubits and ancillary qubits after performing MLR, as shown in Fig.~\ref{fig:DQLR}. The LeakageISWAP gate executes an ISWAP gate that acts non-trivially on the $\{\ket{11},\ket{20}\}$ subspace (rather than the usual $\{\ket{10},\ket{01}\}$ subspace for ISWAP), and removes leakage in data qubits by converting the $\ket{02}$ state to $\ket{11}$. Finally, MLR is performed again on ancillary qubits.

\section{Results}\label{sec:result}
Here we present our simulation results for the codes and leakage removal strategies described in Sec. \ref{sec:problem_setting} that we have calculated using the MPS method described in Sec. \ref{sec:tensor_network_methods}. The tensor-network algorithms are implemented using the Python library TensorNetwork~\cite{roberts_TensorNetwork_2019}. The numerical calculation runs on a machine with Intel Xeon Platinum 9242 system with 96 threads. Each data point on the plots is obtained by averaging ten thousand samples. 

\subsection{Bond dimension}

\begin{figure}[t]
    \centering
    \includegraphics[width=0.8\linewidth]{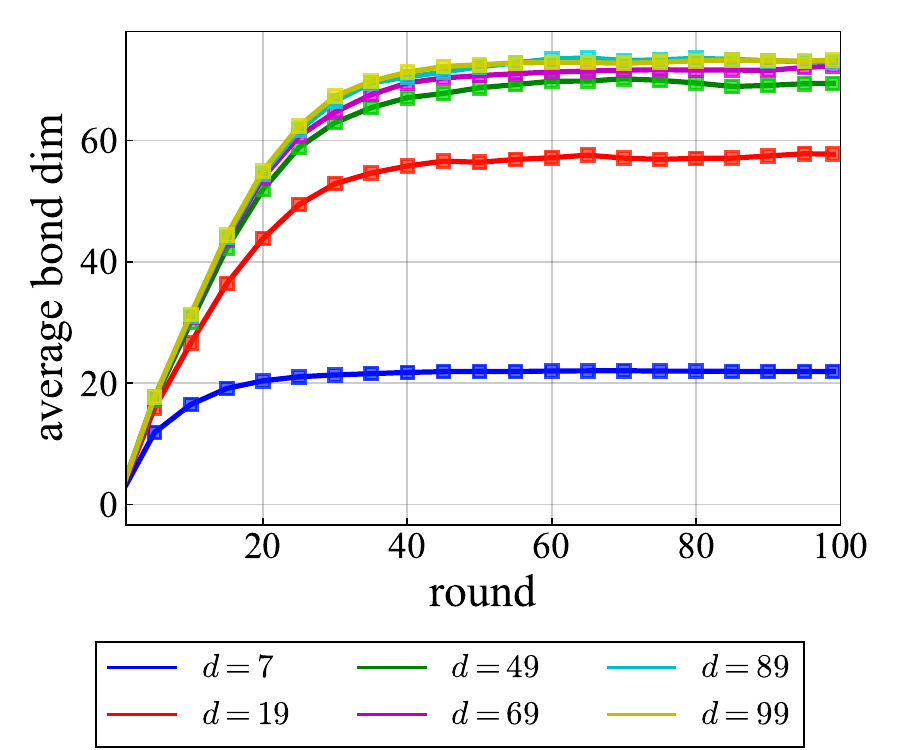}
    \caption{The bond dimension of the MPS at the end of each syndrome measurement round for the 1D repetition code simulation. The bond dimension is truncated at each step to ensure the truncation error does not exceed a threshold. The parameters are set as $T=100$, $\theta=0.1\pi$, $\gamma=0.0$, $\theta_{\mathrm{spread}}=0.3\pi$ and no leakage removal strategy is applied, which requires the largest average bond dimensions in Fig.~\ref{fig:bond_dim_ave}. We repeat syndrome measurements $99$ times.}
    \label{fig:bond_dim_infidelity}
\end{figure}

In order for simulations with MPS to be efficient, it is essential that the bond dimension required for accurate simulation does not grow too quickly in time or with increasing system size. In our simulations, we adopt an approach where the bond dimension is chosen dynamically in order for the truncation error, which is the 2-norm of the vector of truncated singular values, of each truncation step to be below some threshold value. In all of our calculations, we have chosen this threshold to be $10^{-6}$ for the 1D repetition code and $10^{-4}$ for the thin surface code, which we have confirmed is sufficient to accurately calculate the logical error rate in the parameter regions studied.

We present the average bond dimension in the simulation of the 1D repetition code of $d=99$ for the various noise parameters in Fig.~\ref{fig:bond_dim_ave}. As shown in these figures, we find that, generally, a small bond dimension is sufficient for accurate simulations across all noise parameters and leakage removal strategies. 
This implies that the amount of entanglement generated in the error correction procedure for both incoherent and coherent noise is small, and therefore can be captured accurately using the MPS simulator.

The bond dimension required tends to increase as the coherent component of the leakage is increased, and decreases with increasing leakage removal. The latter trend can be understood due to the fact that as the leakage population increases, more entanglement can be generated.

Fig.~\ref{fig:bond_dim_infidelity} shows the bond dimensions during the process in the most computationally intensive noise setting.
The required bond dimensions increase linearly at initial rounds and saturate at a certain point. 
We also find that in the limit of many rounds of error correction, the required bond dimension is constant with respect to $d$, with the exception of finite-size effects for systems with $d\lesssim50$.
From this, it can be inferred that this system exhibits an area law~\cite{eisert_Area_2010,cirac_Matrix_2021}, that is, the entanglement entropy is proportional to the boundary of the two regions (a constant amount of entanglement in the 1D case). This is in contrast to the general case, where exponential bond dimensions are required to represent the quantum state of a $n$ qutrit 1D brickwork circuit with a depth $n$. These results all indicate that our method is scalable and accurate for the codes and noise models under consideration.

\begin{figure}[t]
    \centering
    \includegraphics[width=\linewidth]{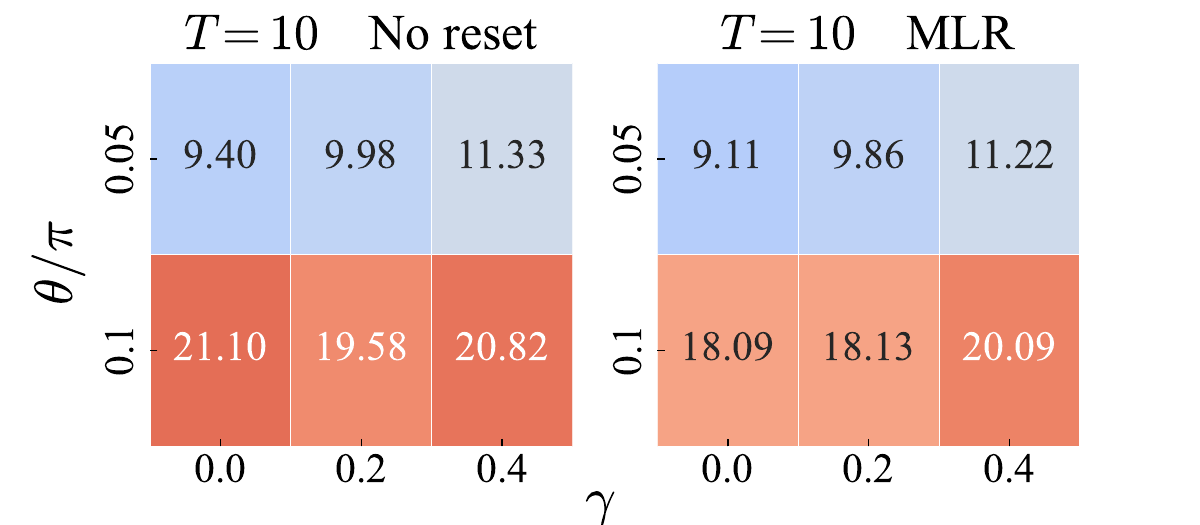}
    \caption{The average MPS bond dimension required to reach a truncation error of $10^{-4}$ when simulating the $3\times 7$ surface code for varying leakage rotation angle $\theta$, coupling strength to bath $\gamma$ and leakage removal strategy. The average MPS bond dimension is computed by collecting all the bond dimensions at the end of each round and taking their average. The number of rounds is $7$.}
    \label{fig:3xn_bond_dim_ave}
\end{figure}

Finally, we plotted the average bond dimension for the $3\times 7$ thin surface code in Fig.~\ref{fig:3xn_bond_dim_ave}, which follows almost the same trend as the 1D repetition code. We note, however, that value is not saturated because of the smaller number of rounds, and the simulation of the $3\times d$ surface code is more costly than that of the 1D repetition code.
%, due to the application of long-range MPO.

\subsection{Exact vs twirling approximation}
\begin{figure}
    \centering
    \includegraphics[width=0.85\linewidth]{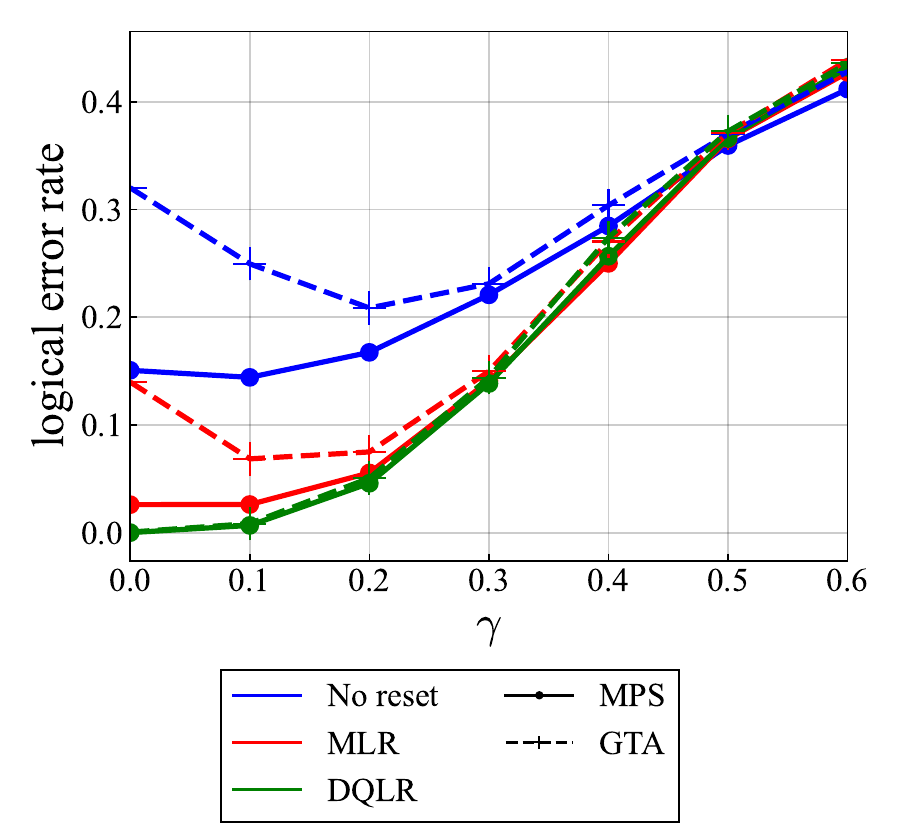}
    \caption{The logical error rate of the 1D repetition code for the coherent noise (solid) and approximated incoherent one (dashed) is plotted as a function of coupling strength to bath $\gamma$. The code distance, temperature, and rotation angle are fixed as $d=19$, $T=10$, and $\theta=0.1\pi$. Three reset strategies, no reset, MLR, and DQLR are plotted in different colors. }
    \label{fig:coherent_incoherent}
\end{figure}
We use our simulation method to examine the difference between coherent and incoherent leakage processes. For this purpose, we consider a well-known method to approximate coherent noise as incoherent noise, the General Twirling Approximation (GTA)~\cite{acharya_Suppressing_2023}. Since our method allows simulations of both coherent and incoherent noise processes, we use it to compare the effect of a coherent leakage noise process to the GTA approximation of it. 

In this test, for simplicity, we assume that there is no leakage spreading, that is, $\theta_\mathrm{spread}=0$. In our model for leakage, which has a single leaked state $\ket2$, we define the state leakage $L$ of a 1-qubit state $\rho$ as 
\begin{equation}
    L(\rho)=\braket{2|\rho|2}\,.
\end{equation}
Next, we define the leakage rate $L_1$ and seepage rate $L_2$~\cite{wood_quantification_2018,wu_leakage_2023} for a given noise map $\mathcal{E}$ as
\begin{align}
    &L_1(\mathcal{E})=L\left(\mathcal{E}\left(\frac{1}{2}\left(\ket0\bra0+\ket1\bra1\right)\right)\right) \\
    &L_2(\mathcal{E}) = 1 - L\left(\mathcal{E}\left(\ket2\bra2)\right)\right)\,.
\end{align}
These rates represent the probability of moving from the computational subspace to the leakage subspace and vice versa. We use them to define a GTA approximated quantum channel $\mathcal{E}'$ to a coherent control leakage unitary $U$ as follows:
\begin{align}
    &\mathcal{E}'=(1-L_1)\mathcal{I}_1+L_1D_{21}+L_2D_{12}+(1-L_2)\mathcal{I}_2 \\
    &D_{21}(\rho)=\mathrm{Tr}[\mathbbm{1}_1\rho]\ket2\bra2 \\
    &D_{12}(\rho)=\mathrm{Tr}[\mathbbm{1}_2\rho]\left(\frac{1}{2}(\ket0\bra0+\ket1\bra1)\right) \\
    &\mathcal{I}_1(\rho)=\Tilde{U}(\mathbbm{1}_1\rho\mathbbm{1}_1)\Tilde{U}^\dagger \\
    &\mathcal{I}_2(\rho)=\mathbbm{1}_2\rho\mathbbm{1}_2
\end{align}
where $\mathbbm{1}_1$ and $\mathbbm{1}_2$ are the projectors onto the computational and leakage subspace, respectively, and  $\Tilde{U}=\mathbbm{1}_1U\mathbbm{1}_1$. Finally, $\Tilde{U}$ is replaced as a stochastic incoherent noise by applying the standard Pauli Twirling Approximation.

\begin{figure*}
    \includegraphics[width=\textwidth]{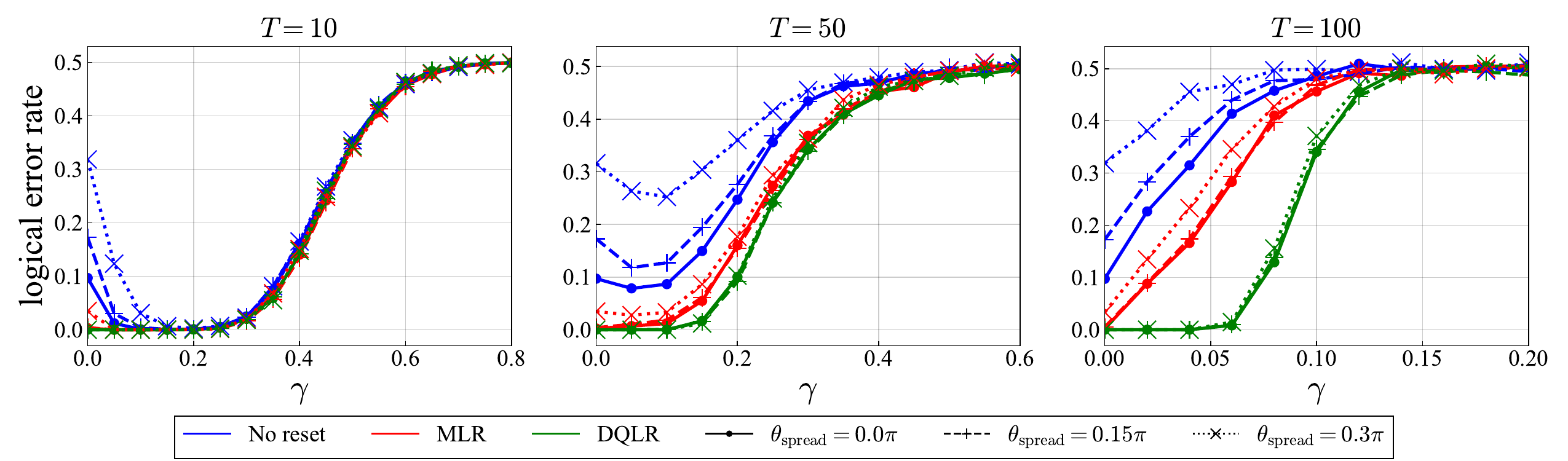}
    \caption{The logical error rate of the 1D repetition code is plotted as a function of coupling strength to bath $\gamma$ for various temperatures and $d=49$, $\theta=0.05\pi$. Three reset strategies are plotted in different colors, and different $\theta_\mathrm{spread}$ values are plotted by different linestyles and markers.}
    \label{fig:high_temp}
\end{figure*}
\begin{figure*}
    \centering
    \includegraphics[width=\textwidth]{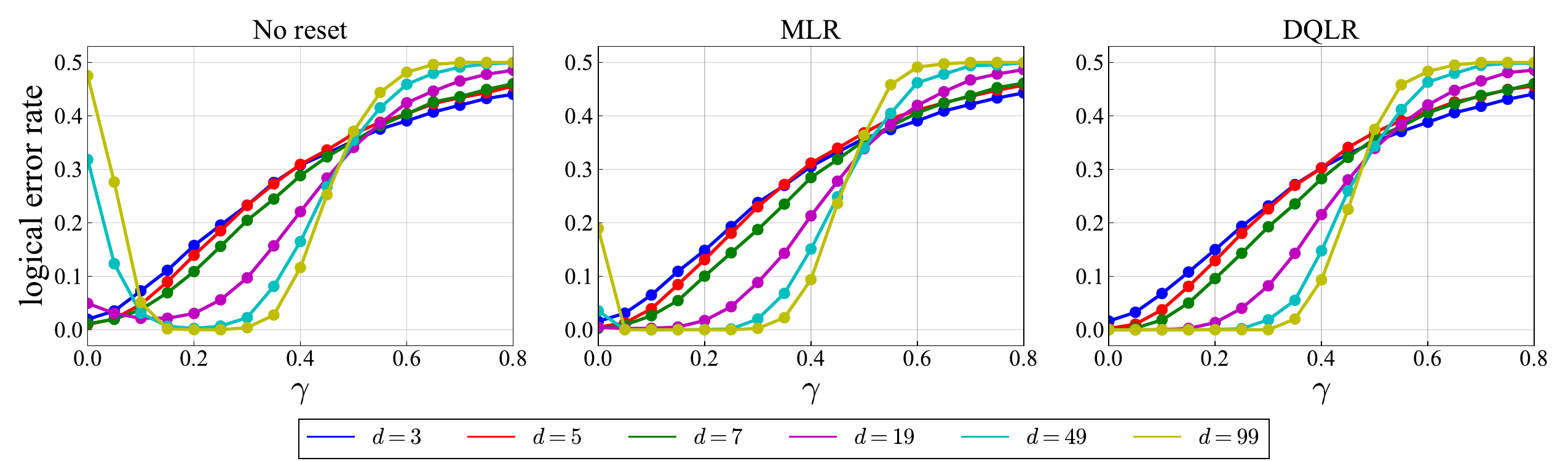}
    \caption{The logical error rate of the 1D repetition code for the three leakage removal strategies as a function of $\gamma$ for various code distances. Parameters are set as $T=10$, $\theta=0.05\pi$ and $\theta_\mathrm{spread}=0.3\pi$.}
    \label{fig:threshold_mlr_noreset_gamma}
\end{figure*}

Fig.~\ref{fig:coherent_incoherent} shows the difference in logical error rates between exact and GTA noise. As can be seen from the figure, introducing an approximation can overestimate the logical error rate by a large amount (by more than a factor of three in the MLR case). These phenomena, where twirling approximations over-predict the logical error rate, have also been observed in earlier studies~\cite{tomita_Lowdistance_2014,katabarwa_Logical_2015,katabarwa_dynamical_2017a}. 
This indicates that the introduction of approximations, like GTA, could potentially lead to misinterpretations of important properties such as inherent error tolerance and thresholds that the code possesses. Moreover, these results demonstrate that the simplified metrics for leakage, such as the leakage and seepage rates, do not necessarily fully capture the actual impact of leakage on the logical error rate.

%The reasons why this leads to overestimation rather than underestimation, and the types of noise for which this becomes more pronounced, remain topics for future research.

\subsection{Comparison of leakage processes and removal strategies}

We conducted numerical experiments to investigate the effect of leakage by computing logical error rates as leakage removal strategies and the parameters of the noise model were varied. 

There are three primary causes of leakage in the noise model: thermal excitation, control leakage, and leakage spreading. In high-temperature regions, the influence of thermal excitation appears strongly, as indicated in Eq.~\eqref{eq:ADTE}, which increases the leakage population. Fig.~\ref{fig:high_temp} shows the logical error rate of the 1D repetition code as a function of $\gamma$, the coupling strength to the bath, as the temperature changes. When adopting less effective leakage removal strategies such as no reset or MLR, which cannot perfectly remove leakage, we observed that the logical error rate significantly increases in the high-temperature region. Additionally, including leakage spreading further deteriorates the performance. In the case of DQLR, the presence or absence of leakage spreading hardly affects the results, indicating that it can nearly ideally eliminate leakage on qubits.

\begin{figure*}
    \centering
    \includegraphics[width=\textwidth]{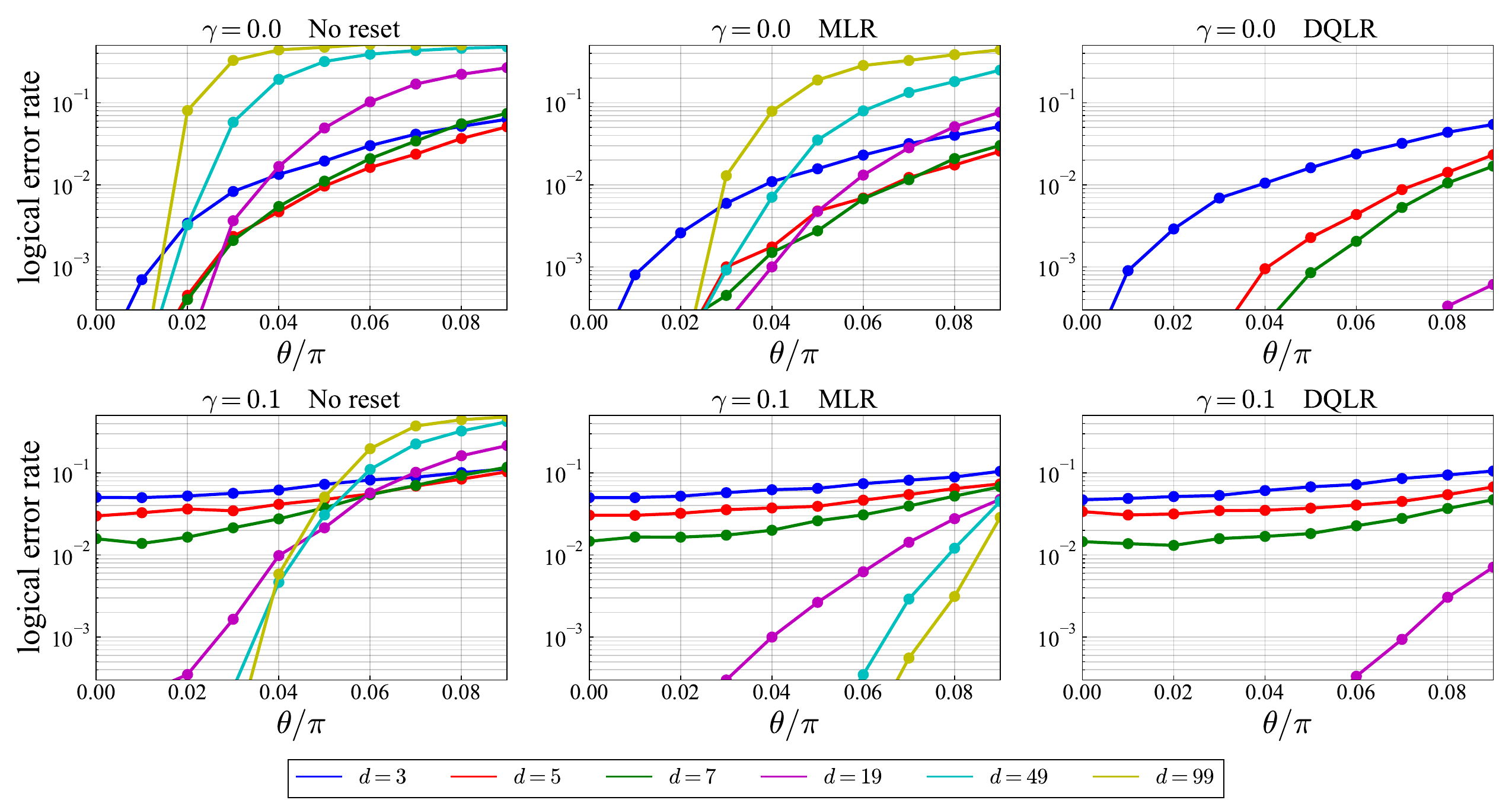}
    \caption{The logical error rate of the 1D repetition code for the three leakage removal strategies as a function of rotation strength $\theta$ for various code distances. Parameters are set as $T=10$ and $\theta_\mathrm{spread}=0.3\pi$. Upper: the results with $\gamma=0$, meaning there are no amplitude damping errors. Lower: the results with $\gamma=0.1$. The logical error rate is calculated based on the results from several thousand samples, so only points above $3\times 10^{-4}$ are plotted for accuracy. Especially, for DQLR, with $d=99$ and $49$, exactly zero logical errors were detected during the simulations.}
    \label{fig:threshold_mlr_noreset_theta}
\end{figure*}

When the temperature is low, leakage is primarily introduced by over-rotation, leading to non-monotonic behavior in the logical error rate. Fig.~\ref{fig:threshold_mlr_noreset_gamma} shows the logical error probability for the three leakage removal strategies in a low-temperature environment for various code distances. For large code distance $d$, in the small $\gamma$ region, the logical error rate sharply increases as $\gamma$ decreases, when we use either no reset or MLR strategy. This trend becomes more pronounced as the code distance $d$ increases. This is because when coherent leakage is strong, the coupling to the thermal bath induces amplitude damping noise, which effectively reduces the leakage population. If the amplitude damping is weak, i.e., when $\gamma$ is small, the leaked state persists and further spreads to its neighbors, fundamentally destroying the logical state composed of many physical qubits. It should be noted that these unusual phenomena cannot be obtained from simulations of small-scale systems.

We also plotted the logical error rate as a function of over-rotation angle $\theta$ in Fig.~\ref{fig:threshold_mlr_noreset_theta}. When $\gamma=0$, these plots show the effect of leakage purely due to fast control.
One can see that for fast-control leakage, the effect on the logical error rate becomes significant for larger system sizes. 
On the other hand, DQLR is very effective at eliminating leakage caused by control leakage and lowering the logical error rate, especially for these large sizes. 
For small $\gamma$ one can observe competition between leakage caused by fast control and amplitude damping that tends to return leaked states to the qubit subspace. As seen above, amplitude damping also tends to reduce the effect of fast-control leakage on larger systems. 

\begin{figure*}
    \centering
    \includegraphics[width=0.7\textwidth]{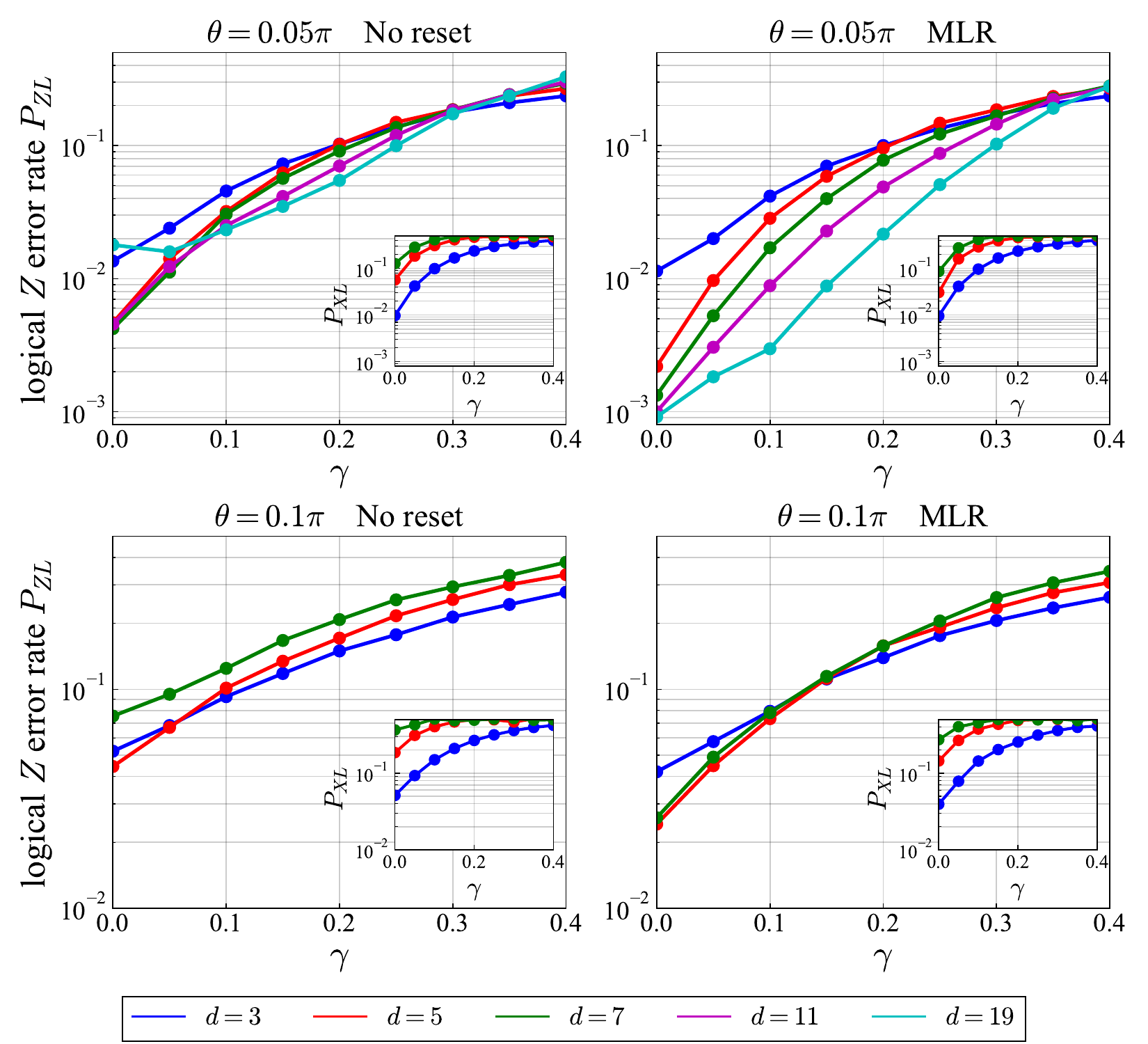}
    \caption{The logical Z error rate $P_{ZL}$ of the $3\times d$ surface code for the two leakage removal strategies and rotation angles $\theta=0.05\pi,0.1\pi$ as a function of coupling strength to bath $\gamma$ for various $Z$-distance $d$. Parameters are set as $T=10$ and $\theta_{\mathrm{spread}}=0$. The inset is the logical X error rate $P_{XL}$ on the same setting.}
    \label{fig:3xn_threshold_mlr_noreset_gamma}
\end{figure*}

Lastly, we discuss the results of the MPS simulation for the $3\times d$ thin surface code. Due to the simulation cost and the complexity of the procedure, we didn't consider leakage spreading, and only No reset and MLR strategies were adopted for the leakage removal. Fig.~\ref{fig:3xn_threshold_mlr_noreset_gamma} shows the logical error rate against $\gamma$ for various $Z$-distance $d$ when the amount of control leakage and the leakage removal strategies are changed. The logical $Z$ error rates broadly follow the same trend as the 1D repetition code, where a larger $\theta$ results in more leakage and a higher logical error rate, but these can be suppressed by employing better strategies. Additionally, we also observed a reversal phenomenon where the logical error rate becomes larger when $\gamma$ is small and $d$ is large, which results in the disappearance of the threshold. We anticipate that such phenomena would also be observable in standard $d\times d$ surface codes.
We plotted the logical $X$ error rates in the inset of Fig.~\ref{fig:3xn_threshold_mlr_noreset_gamma}. As we expected, the logical $X$ error rates increase as code distance increases, but the leakage removal strategy is still effective in suppressing the noise effect.

In our study, we just use one CPU node to execute the simulation.
For the $d\geq 5$ simulation, the cost of performing SVD becomes a crucial bottleneck of the computation time, so massive parallelization on cluster machines or fast SVD execution using GPUs is required.
More realistic problem settings and large-scale simulations are left as topics for future research.

\section{Conclusion}\label{sec:conclusion}
In this study, we propose a method for simulating leakage processes in quantum error correcting codes of up to several hundred qudits using the MPS method. This approach enables us to investigate the severe impact of coherent leakage errors and thermal noise on QEC codes without simplifications such as the general twirling approximation.

We demonstrate that applying such approximations can lead to substantially different performance estimates compared to the original noise model. For instance, we observe that an incoherent approximation to a coherent leakage noise model can significantly overestimate the logical error rate. The result of the numerical calculation emphasizes that the simplified indicators for leakage, like the leakage and seepage rates, fail to fully encapsulate the actual impact of leakage on the logical error rate.
From a practical perspective, we also highlight the importance of various leakage removal strategies. We show that the effect of these strategies varies drastically depending on the specific type of leakage process. 

Furthermore, this study also introduces a new application of quantum circuit simulations using tensor networks. In recent years, tensor-network methods for quantum computation have greatly advanced and have been employed for various tasks, such as the verification of quantum advantage~\cite{zhou_What_2020,gray_Hyperoptimized_2021,yong_Closing_2021,pan_Solving_2021,huang_Efficient_2021,oh_Tensor_2023,tindall_Efficient_2023} and for benchmarking and pre-training in variational quantum algorithms~\cite{stoudenmire_Supervised_2016,han_Unsupervised_2018,huggins_Quantum_2019,dborin_Matrix_2021}. 
To effectively apply tensor networks to a problem, it is essential to utilize geometric and entanglement structures of the system. In this study, we exploited a low-entanglement structure to discover non-trivial behavior in the processes of 1D quantum error correction codes under a leakage noise model. 

Looking forward, it is expected that other tensor network algorithms will also be utilized for evaluating the memory performance beyond 1D and quasi-1D codes. To investigate codes with higher dimensional structure, more sophisticated tensor-network algorithms will likely be required, such as PEPS and isoTNS~\cite{evenbly_Gauge_2018,zaletel_Isometric_2020}.

\section*{Acknowledgements}
%\begin{acknowledgments}
H. Manabe acknowledges K. Harada for discussion about tensor networks. H. Manabe is supported by JSTCOI-NEXT program Grant No.JPMJPF2014.
Y. Suzuki is supported by PRESTO, JST, Grant No.\,JPMJPR1916; MEXT Q-LEAP Grant No.\,JPMXS0120319794 and JPMXS0118068682, JST Moonshot R\&D, Grant No.\,JPMJMS2061.
A. Darmawan was supported by JST, PRESTO Grant Number JPMJPR1917, Japan.
%\end{acknowledgements}

%\bibliographystyle{apsrev4-2}
\bibliography{reference_manabe,reference_suzuki}

\appendix

\section{Quantum circuit simulation with MPS}\label{app:Quantum_ciricuit_simulation_with_MPS}
We outline how the evolution of the many-qudit quantum state over many rounds of quantum error correction is captured in the MPS formalism. First, we express the initial state as an MPS. In our simulation settings of a noisy one-dimensional quantum circuit, single-qubit gates, single-qubit CPTP maps, and two-qubit gates are applied during error correction, all of which correspond to simple updates to the tensors of the MPS representing the state. The application of a single-qubit gate $M_i$ at site $i$ simply updates the corresponding tensor as follows:
\begin{equation}
    A_{\alpha_{i-1}\alpha_i}^{[i]s_i}\leftarrow \sum_{s_i'}M_{s_i'}^{s_i}A_{\alpha_{i-1}\alpha_i}^{[i]s_i'}
\end{equation}

To apply a two qubit gate to a pair of neighboring qubits $i$ and $j$ in the MPS, we simply contract the corresponding physical indices of the MPS to a two-qubit Matrix Product Operator (MPO) representing the gate 
\begin{equation}
    M_{s_i's_j'}^{s_is_j}=\sum_{\alpha}M_{\alpha}^{[i]s_is'_i}M_{\alpha}^{[j]s_js_j'}\,,
\end{equation}
which can be obtained by applying the singular value decomposition to the tensor representing the 2-qubit gate. 
We can describe two-qubit gates on non-neighbouring qudits $i$ and $j$ ($i<j$) by including the additional intermediate tensors 
\begin{equation}
    M_{\alpha_{l-1}\alpha_l}^{[l]s_ls_l'}=\delta_{\alpha_{l-1},\alpha_l}\delta^{s_l,s_l'}  \quad (i<l<j)
\end{equation}
in the MPO on all sites between $i$ and $j$.

When an MPO is applied to the MPS, its bond dimension increases correspondingly. In order to suppress the bond dimension, an appropriate approximation using the canonical form is necessary. The canonical form is defined as follows, a constrained MPS format:
\begin{align}
    \sum_{s_l,\alpha_{l-1}}A_{\alpha_{l-1}\alpha_l}^{[l]s_l}A_{\alpha_{l-1}\alpha_l'}^{[l]s_i\dagger}=\delta_{\alpha_l\alpha_l'} \quad (l<k) \\
    \sum_{s_l,\alpha_l}A_{\alpha_{l-1}\alpha_l}^{[l]s_l}A_{\alpha_{l-1}'\alpha_l}^{[l]s_l\dagger}=\delta_{\alpha_{l-1}\alpha_{l-1}'} \quad (l>k)
\end{align}
The tensor at position $k$ is referred to as the top tensor whereas all other tensors except for the top tensor are isometry. When the two-qubit gate acts on the $i$th and $j$th qubits, the top tensor is first moved to the $i$-th position by repeatedly performing singular value decomposition (SVD). Then, the MPO tensor is applied in order from the $i$-th. After the MPS and MPO tensors at the $l$-th and $l+1$-th positions are contracted together, this tensor $A_{\alpha_{l-1}\alpha_{l+1}\alpha_{l+1}'}^{[l,l+1]s_l,s_{l+1}}$ is split into two tensors using singular value decomposition:
\begin{equation}
    A_{\alpha_{l-1}\alpha_{l+1}\alpha_{l+1}'}^{[l,l+1]s_l,s_{l+1}} = \sum_{\alpha_l}U_{\alpha_{l-1}\alpha_{l}}^{s_l}S_{\alpha_l\alpha_l}V_{\alpha_l\alpha_{l+1}\alpha_{l+1}'}^{s_{l+1}\dagger}
\end{equation}
At this time, an approximation is performed by truncating the small singular values. In this study, the threshold of the truncation error is set to $10^{-4}$ or $10^{-6}$ to achieve a balance between low bond dimensions and high fidelity. Finally, the MPS is returned to the canonical form by setting tensors as follows:
\begin{align}
    &A_{\alpha_{l-1}\alpha_{l}}^{[l]s_l} = U_{\alpha_{l-1}\alpha_{l}}^{s_l} \\
    &A_{\alpha_{l}\alpha_{l+1}\alpha'_{l+1}}^{[l+1]s_{l+1}} = \sum_{\alpha_l'}S_{\alpha_l\alpha_l'}V_{\alpha_l'\alpha_{l+1}\alpha_{l+1}'}^{s_{l+1}\dagger}
\end{align}
We repeat this process until the $j$-th tensor is contracted.

Noise such as amplitude damping and thermal excitation are represented as single-qubit CPTP maps. To perform pure-state simulations for these processes, we sample the Kraus operators $K_i$ according to the probabilities
\begin{equation}
    p_i=\mathrm{Tr}(K_i\ket\psi\bra\psi K_i^\dagger)\,,
\end{equation}
which can be calculated efficiently using the MPS ansatz. The sampled Kraus operator can be applied similarly to a single-qubit gate. Projective measurements can be simulated in the same manner.

\end{document}